# The controversial pen of Edwin Holmes

Jeremy Shears

**Abstract**

Edwin Alfred Holmes (1839 -1919) is best remembered for his discovery of a bright comet in 1892, now known as Comet 17P/Holmes. An amateur astronomer and authority on optics, he was an original member of the BAA and contributed to its *Journal* and meetings for many years. As a prolific writer of letters to the *English Mechanic,* he developed a reputation for his controversial and acerbic penmanship.

**Introduction**

The name of Edwin Alfred Holmes (Figure 1) will forever be associated with the eponymous comet that he discovered, quite by chance, just before midnight on 1892 November 6, whilst trying to locate the Andromeda Nebula, M31. He had observed the Nebula regularly ever since the appearance of a bright new star in 1885, now known to have been a supernova (1). On pointing his telescope in the direction of the Nebula he placed his eye to the eyepiece and was shocked by its unusual appearance. Holmes said (2) he "called out involuntarily, 'What is the matter'? 'There is something strange here.' My wife heard me and thought something had happened to the instrument and came to see." He quickly realised that it wasn't the Nebula, but a bright comet, saying to his wife "This is coming end on, and will be a big fellow".

Realising the importance of his discovery Holmes notified the Royal Observatory at Greenwich. The report was initially received with scepticism, for perhaps the amateur observer had mistaken it for the Andromeda Nebula (Holmes had omitted to tell them he had established it to be a different object). Nevertheless the discovery was confirmed as a new comet on the evening of November 7. It transpired that it had passed perihelion nearly five months earlier, but at the time of discovery was undergoing a massive outburst in apparent brightness, bringing it to naked-eye visibility (3). A drawing made during the outburst by H.F. Newall (1857-1944) at the Cambridge Observatory is shown in Figure 2. It began to fade in the second half of November and a second outburst occurred in mid-January 1893. Holmes was awarded the Donohoe comet medal of the Astronomical Society of the Pacific in recognition of his discovery.

The comet, officially known as 17P/Holmes, is now understood to be a member of the Jupiter family of short-period comets, with a period of about 7 years. Although many of its returns in the intervening years were missed (4), observers around the world were treated to a *megaburst* of 17P/Holmes during its 2007 apparition when it reached naked eye visibility and attracted much public and media attention. The CCD image of the 2007 outburst by Richard Miles in Figure 3 shows the huge spherical dust-coma and is remarkably similar to Newall's 1892 sketch (Figure 2).

Whilst this might have been Holmes's only major discovery, he was nonetheless a well-known amateur astronomer, through his contributions to the BAA, of which he was an original member and to whose *Journal* he submitted several papers, and through hundreds of letters he wrote to the *English Mechanic* concerning the subjects of astronomy and microscopy. The *English Mechanic and World of Science*, to give it its full title, was published weekly from 1865 to well into the twentieth century. It was famous for its articles concerning all branches of science, engineering and technology and was equally well-known for its letters pages in which readers would pose questions, offer advice on various topics as well as describe their own experiments and observations. It was dearly loved by its readership who generally referred to it simply as "*Ours*". Given the firmly held views of many readers the correspondence often became heated and even vitriolic. If the reader thinks that such behaviour, which can be encountered regularly even on astronomical discussion groups in the form of internet trolls, is a feature of the modern era, then they should look at some of the exchanges within the pages of the *English Mechanic*! It was through this organ that Holmes developed a reputation for controversy through his sometimes unguarded, and often vitriolic, comments on various topics to do with astronomy and optics, and his criticism of the views of other correspondents, some of which became quite personal in nature. On several occasions he was publically rebuked by the Editor for overstepping the mark. On the other hand, whilst Holmes's letters could be acerbic, contemporary descriptions of those who knew him well indicated that he was in reality a mild-mannered and considerate man. One person who knew him well was Arthur Mee (1860-1926) of Glamorgan (5), who had exchanged letters with Holmes in the *English Mechanic* and had fallen foul of Holmes's pen. Mee reflected on Holmes's character (albeit posthumously) thus: (6)

"I…found him a most kindly, genial and helpful man. He was never so happy, I am sure, as when assisting others over stiles. Critical himself, Mr. Holmes harboured no malice"

So who was Edwin Holmes and what were some of the controversies that unfolded within the pages of the *English Mechanic* (henceforth "*EM*") and elsewhere?

**Biographical sketch**

Holmes was born in Sheffield in 1839, eventually moving to London where he was occupied as a glass merchant and glass cutter. He married Selina Stevens of Shoreditch in 1864 and the couple had a son, Ernest, in 1875. For much of the time they lived at "Telescope House", Hornsey Rise, Islington, where he built a small observatory (7) (Figure 4). Selina passed away in 1907 and Holmes later moved to Tottenham where he died on 1919 January 21.

Holmes used a variety of telescopes over the years. These were mainly reflectors as his experience showed that a well made reflector could outperform a refractor of similar cost. As we shall see later, his forthright views on telescope performance which he expressed in dozens of letters in the *EM*, led him into a number of disputes. By the 1890s Holmes was using a 9-inch (23 cm) Newtonian on an altazimuth mount (8) (Figure 5) and later in the decade he had a 12¼-inch (31 cm) reflector of very similar design (Figure 6) (9).

Holmes's observational tastes were catholic and included the Moon, planets, double stars and what would today be called deep-sky objects. He contributed to the work of several of the BAA Observing Sections, but interestingly he was not a member of the Comet Section at the time of his comet discovery, although he joined shortly afterwards and reported observations of other comets over the years.

**Debates about optics and telescope performance**

It was through his professional activity in the glass trade that Holmes developed an interest in optics, which led quickly to his active pursuit of both microscopy and astronomy. He began contributing to the *EM* in 1867. His initial letters were on the subject of trades unions, but his first contribution on optics was in 1871 January when he described how he a constructed a "cheap micro-polariscope" (10). The following month he became involved in a discussion on silver-on-glass mirrors. Controversy first raised its head in 1874. Holmes wrote to extol the quality of telescopes made by a certain W.J. Lancashire: "standing on the beach at West Cowes, [I] could see the ropes of the shipping at Hurst Castle, 18 miles away" (11). When John Hampden (1819-1891) questioned the veracity Holmes's observation (12), Holmes took umbrage: "The extremely insulting way he [Hampden] chooses to express himself in regard to a matter of which he knows nothing, and to a person of whom he knows less, would have made me treat him with the contempt he merits, but that other correspondents might accept his rash contradiction as a correct view of the matter. He flatly calls me a liar…" (13). Hampden was a notorious flat-earther and was considered a *bête noir* in the *EM* for trying to promote his flat earth theories and for his outrageous hectoring of the Editor over a period of many years. He was only really using his reply to Holmes further to expound his "flattist" views. In Hampden's opinion the fact that Holmes could see such a distant object was further evidence for the earth being flat. Quite why Holmes decided to respond to somebody who was widely recognised as a charlatan is a mystery (14).

The next major controversy occurred in 1880 and the subject was one close to Holmes's professional experience: the etching of glass with acids. What started out as a discussion on the advantages of various types of acid, and the dangers of hydrofluoric acid, soon deteriorated into a bad tempered exchange between Holmes and Alfred Henry Allen (1846 – 1904; Figure 7), an expert in chemical sciences from Sheffield (15). It perhaps didn't help when Allen suggested that Holmes might be suffering from a "mental and observational incapacity" which could be influencing his views. Not merely content with using the pages of the *EM* to debate the matter, Allen reported that Holmes also resorted to sending him "a grossly abusive letter through the medium of the post. If Mr. Holmes continues to favour me with his communications, as long as he pays the postage I shall take them in; but I must decline to continue this correspondence, as I have neither time nor inclination to reply to personalities designed to obscure the real questions at issue" (16). Allen forwarded the offending letter to the Editor who duly noted that "Mr. Holmes has committed an indiscretion in writing to Mr. Allen at all. We prefer to say nothing of the nature of his communication".

Apparently undeterred, Holmes threw himself into other, more good-humoured correspondence on the merits of refractors versus reflectors, generally supporting the latter, assuming they were well made, since with a given sum of money it was possible to purchase a much larger reflector. He narrowly escaped offending the telescope maker Henry Wray in saying "I think however valuable the opinion of a maker may be, he is necessarily prejudiced in favour of his own productions. Those who use telescopes of both kinds are the best judges of their comparative merits" (17). Whilst this may be true, it was slightly tactless, and fortunately overlooked by Wray, although it was not untypical of the robust tone of many *EM* correspondents.

Holmes's next sparring partner was W.S. Franks (1851-1935), in 1885, and the subject once again was the relative performance of telescopes. Franks was a highly respected amateur astronomer with a special interest in star colours. His first major publication was *A Catalogue of the Colours of 3890 Stars* which was communicated to the RAS on his behalf by the Reverend T.W. Webb (1807-1885) in 1878. In the 1880's Franks served as Director of the Star Colour Section of the Liverpool Astronomical Society, a role which he would subsequently hold in the BAA from its establishment in 1890 (18). Franks was by all accounts a mild-mannered individual, but he felt compelled to object to some of Holmes's rhetoric: "I shall always be glad to assist him [Holmes] when in my power, the same as anyone else. But I must deprecate strongly the practice of introducing covert insinuations, more or less personal, into scientific discussion; it is both unkind and ungentlemanly, and can do no good whatever", further accusing Holmes of behaving in a "cold-blooded, cynical manner, suggestive of gall and wormwood, which only wounds" (19). Holmes replied saying "I feel called on to deny that I was at all wanting in courtesy to Mr. Franks except that I was uncourteous enough to differ from him, and to prove I was right" (20). Franks ended the exchange there by saying "I decline to hold any further communication whatever with that gentleman" (21).

In the ensuing correspondence about telescope performance, Holmes had also managed to cast aspersions on the judgement of the renowned mirror maker George Calver (1834-1927), who went on to say "I hope Mr. Holmes does not expect me to reply to his cavilling and jeering comments on my letter. When he ceases to ape at being witty in order to make others look foolish he may be worth noticing….Mr. H. has well earned the name of a 'carping hypercritic' " (22). Once again Holmes felt compelled to respond "I protest against the language by Mr. Calver in reply to civil questions…. It is too much, when one is only asking for information, for the Editor to admit such personalities, and then cut out of the letters of the person attacked every word that indicates his opinion they are unjust" (23). The Editor, clearly exasperated, put a stop to the affair: "It is always 'the Editor' who is in fault, according to Mr. Holmes. Our correspondent has scarcely ever come to words with his fellow-correspondents without blaming us for it. Besides the above, he favours us with a private letter, in which other correspondents are mentioned as having apologised to Mr. Holmes for their remarks in these columns. If so, Mr. Holmes may be left to enjoy what he dearly loves—the last word.—Ed." In spite of all this, Holmes must have respected Calver's abilities as an optician for he got Calver to refigure the mirror of his 12¼-inch Newtonian some years later.

For many years, one of the most active and robust *EM* correspondents, almost to the point of libel on occasion, wrote under the name "*A Fellow of the Royal Astronomical Society*" or sometimes simply "*FRAS*". The use of pseudonyms and abbreviations in *EM* was common practise and in most cases was not intended deliberately to hide the identity of the person, in this case Captain William Noble (1828-1904), who served as the first president of the BAA upon its formation in 1890 (whilst Holmes normally signed his letters "Edwin Holmes", he sometimes used "Alfred", or "A."). Noble's strong character, a man known for his "sturdy independence" (24), is perhaps exemplified by his BAA presidential portrait in which he is pictured wielding a large shotgun (25). Although Holmes did not have any major disagreements with Noble, he certainly seized the opportunity of pointing out a glaring error in one of Noble's letters where he incorrectly gave the weight of a sovereign coin, going on to say: "There is some slip which, if he allows to pass unexplained, we shall have adduced as evidence of his ignorance" (26). Noble, perhaps wisely, chose not to rise to Holmes's bait.

**The great "refractors versus reflectors" debate**

It was the polite Arthur Mee, who as we saw earlier ultimately held great admiration for Holmes, with whom Holmes next crossed swords. Mee had written a letter asserting that in his experience refractors were better than reflectors for observing the sun. It was therefore no surprise that Homes, with a predilection for reflectors, should advocate a different perspective giving examples of how he had had better views with a 6-inch Newtonian than with a 4-inch achromatic refractor, suggesting that tube currents in reflectors could cause problems, but their effect could largely be eliminated through proper design of the telescope tube. However, as was so often the case, Holmes didn't stick to stating the facts, as he saw them, and instead moved into innuendo: "I do not expect the 'many' to come forward [to support my views]. The ordinary astronomical amateur, however free in private conversation, is not fond of public utterances, especially when he has to differ from eminent authorities and be sat upon. It is only a few erratic individuals like myself who risk exposing themselves to the sarcasm which runs through Mr. Mee's letter. Mr. Mee is an F.R.A.S., and a very eminent astronomer. I am no more than a mere star-gazer, and am willing to concede that I am hardly qualified for membership of the B.A.A." (27). Mee replied in the following week's edition pointing out that "Our friend not only gives a shifty answer himself to my straightforward questions, but his failure to produce so much as one of the 'many' observers who agree with him is an unpleasant set-off against his own indubitably wide experience" (28).

But it was a later sentence in Mee's letter that set another hare running: "There has been too much special pleading for both reflectors and refractors by persons directly or indirectly interested, in their disposal". Of course, Holmes had written dozens of letters which mainly promoted reflectors; what was different this time was that Holmes had recently sold his 9-inch reflector and the purchaser soon afterwards had sold it on to yet another individual. Could Mee have been suggesting that it was in Holmes's personal interest to be promoting reflectors at about the same time as he was selling such an instrument? Whatever was in Mee's mind, Holmes interpreted it as an attack on his honesty and integrity. As if this weren't enough, another pseudonymous writer, the self-styled and anonymous "*Truth*", made a further claim about a telescope which had once belonged to an "amusing and instructive correspondent of 'ours', was soon sold again at a heavy sacrifice by the unfortunate purchaser. A 12¼-in. which replaced it has been vainly offered for resale for months past, a gentleman to whom it was lent on trial rejecting it for a much smaller refractor in less than ten minutes" (29). It was plain for all to see that *Truth* was referring to none other than Edwin Holmes! Naturally enough, Holmes felt compelled to respond to Mee's and *Truth*'s allegations of bad faith, noting that "the statements of 'Truth' are so untruthful, I feel called on to state the facts" (30), which he went on to do, declaring that there was nothing whatsoever to the allegations and that he refused to consider the matter further. Again it was left to the Editor to draw a line under the squabble by inserting after Holmes's letter: "Thank Goodness! How is it that some of our best correspondents have such thin skins, and fancy that the great majority of readers are in the least interested in their personal squabbles?-Ed." It was more than 9 months before Holmes resumed writing to *EM*, an uncharacteristically long gap.

**Holmes's nemesis emerges: James Hunter FRSE, FRAS**

A debate which raged for 6 months during 1898 was whether Newtonian reflectors were superior to Gregorians. This was of course grist to Holmes's mill: he certainly had opinions on the matter, largely in favour of Newtonians, which he was pleased to have the opportunity of sharing! During the exchanges a new combatant emerged who simply signed himself "*H.*" and who was a frequent contributor on many subjects to *EM*. It later transpired that "*H.*" was the well-respected surgeon James Hunter, FRSE, FRAS, a Fellow of the Royal College of Surgeons of Edinburgh and Lecturer in Physiology at the Royal Colleges of Medicine in the same city. Hunter was a keen amateur astronomer with a deep interest in optics (31). His correspondence with Holmes contained much humour, almost to the point that suggests he actually enjoyed teasing and provoking Holmes. One particularly amusing episode was when he playfully linked "our eminent" Holmes with his literary namesake, the popular fictional detective Sherlock Holmes, whose death at the Reichenbach Falls in 1893 the British public were still mourning (32). After 2 months of almost weekly letters back and forth between Holmes and Hunter, the *EM* Editor was clearly becoming vexed: "If either 'H.' or Mr. Holmes wish to continue this discussion, or any other correspondents, they will please stop aggravating each other by these angry personalities. There must be something about this subject akin to rabies!" (33).

Nevertheless, neither Holmes nor Hunter took heed of the Editor's advice and the debate rolled on. Other correspondents joined the fray including "A.S.L." (34) who accused Holmes of talking nonsense – which in turn elicited another string of letters in reply from Holmes. After 6 months, Holmes was finally running out of steam, noting "I am done with Gregorians &c". The following week the last letters from Holmes and Hunter on the subject were published. Hunter couldn't resist a final jibe about "the ludicrous impossibilities [proposed] by a person who signs himself in these pages 'Edwin Holmes' " (35). Holmes simply concluded his letter: "I decline further discussion with 'H.' ". (36).

This was not the last that Holmes would hear from Hunter, but it was Holmes's last *EM* letter for nearly three years, except for a brief note reporting his observations of Nova Persei 1901 in March (37). When he resumed writing it was mainly factual accounts of his observations of double stars and his experiments with spectroscopy and photography, with controversy being given a wide berth.

However, in 1903 Holmes engaged in a new debate with several readers, including the Reverend Charles L. Tweedale, on miracles, near death experiences and religious belief – always a touchstone for controversy. Tweedale was an Anglican minister from Weston in Yorkshire and a leading voice for spiritualism in Britain who was convinced that images of the spirits of deceased people could be captured through the medium of photography. An example of a supposed spirit photograph of Tweedale's late father-in-law is shown in Figure 8. There were some sharp exchanges and whether it was related to this subject or not, the Editor decided to nip in the bud any escalation and chose not to publish one of Holmes's letters, noting in the section headed *Hints to Correspondents*: "EDWIN HOLMES.—Better let it pass. We give effusions of the kind for what they are worth. Sometimes they suggest incidentally trains of useful thought; sometimes they—not unfairly, we must admit—offer targets for ridicule. But attacks only invite replies, which, on the whole, waste space." (38) In the event Holmes's correspondence moved onto other matters such as the nebular hypothesis of the formation of the solar system, the merits of wooden telescope tubes and, perhaps inevitably, the familiar debate which can be summarised as "*refractors vs reflectors*".

**Criticism of the BAA**

As mentioned earlier, Holmes became an original member of the BAA upon its formation in 1890. His contributions to the *Journal* were prolific; his first paper "*On the Visibility of the*

*Disk of Titan*" appeared in the fourth edition of the *Journal* (February 1891) (39) and the last was in 1917 on the dark lanes in the nebula M51 which were visible in photographs (40).

Living in London meant that Holmes was able to attend nearly every BAA meeting, often presenting papers, asking questions and engaging in discussion (41). His presence at meetings was much appreciated for the humour he injected and also for the knowledge that he was able to impart based on his years of practical experience. With one notable exception, there are no accounts of major public disagreements or discord at the meetings, although these would have probably not been recorded anyway, and it appears that his controversial tone was largely reserved for the written word (42). The exception relates to a paper Holmes read at the 1906 April BAA meeting in which he criticised some of E.W. Maunder's (1851-1928) work on the links between solar activity and magnetic storms (43). When Holmes sat down, Maunder took the opportunity to dismantle Holmes's points one by one. His rebuttal was covered in excruciating detail in meeting report published in the *Journal* where it occupied six full pages, two more pages than Holmes's original paper!

It was in 1906 that indications that Holmes was not entirely happy with affairs at the BAA emerged publicly through the pages of the *EM*. Whether it was connected with the Maunder incident or not, the immediate catalyst was a letter from H.P. Hollis (1858-1939) to the *EM* in 1906 November commenting on the recent decline in membership numbers, which had dropped from around 1,200 in 1900 to 1,000 at the time of writing (44). Hollis went on to point out some of the benefits of joining, encouraging all those who had an interest in astronomy to do so. The first response was from the BAA member, A.A. Buss, of Manchester, who voiced some complaints about the content of the *Journal*, especially what he considered to be the excessive level of detail with which its London meetings were covered, the lack of coverage of provincial meetings, and the fact that he wished to see more observational material presented (45). Charles Grover (1842-1921), of the Rousdon Observatory in Dorset, also commented in the same edition, but he was supportive of the Association and the *Journal*'s contents (46). The following week's edition carried a letter from Holmes, largely in response to Buss's criticisms (47). The letter was also generally supportive of the BAA, its meetings and *Journal*, although he agreed there was always room for improvement. He disagreed with Buss about including more material relating to provincial branches of the BAA, stating his "opinion that the establishment of branches was a mistake". Buss, who had been a leading light in the North West Branch of the BAA that operated between 1892 and 1903, took exception to Holmes's comments (48) and what he thought were his London-centric views.

There the matter might have lain if it were not for a contribution from an anonymous "*S.B.*": "Reading between the lines of letters, I infer …….Mr. Holmes is offended at his exclusion from the Council. Let me appeal to both [Holmes and Buss] to forego their not unnatural resentment. We cannot all sit on the Council, and we should not let personal feelings interfere. Mr. Holmes has had his grumble before, but he is in a sad minority. He is not the only one who thinks he deserves a front seat in the synagogue." (49)

"*An Original Member of the BAA*" recognised the imminent danger posed by *S.B.*'s comments: "We often hear of the desirability of pouring oil upon the troubled waters. Apparently your correspondent, 'S. B.,' has heard of the proverb, and has jumped to the conclusion that the same soothing liquid has an equally satisfactory result when poured on a smouldering fire." (50) Much to the relief of readers, after a further exchange of letters between Holmes (who claimed that "Mr. Buss's letter has degenerated into a mere personal attack upon myself" (51)), Buss and some other BAA members, the matter died away. Holmes's name had in fact been put forward for election to Council in 1906, but members were informed at the June meeting that he had withdrawn his nomination (52), contrary to *S.B.*'s assertion that he had been excluded. One wonders whether the withdrawal was in connexion with the public mauling Holmes received at the hands of Maunder, who was not only a Vice-President and prominent Council member, but who was also regarded as one of the Association's leading members, at the April meeting mentioned above.

Shortly after this episode, the performance of Gregorian reflectors was debated once again in the *EM* and Holmes clearly felt compelled to contribute. Holmes's letter (53) was noticed by his arch-protagonist "*H.*", James Hunter, who had not contributed to the *EM* for several years. What annoyed Hunter (54) was that Holmes had laid claim to having been the first to present a formula for calculating the equivalent focus of a compound telescope, which Hunter had in fact published first. Perhaps fearing a repeat of the 6-month long battle between Hunter and Holmes that had taken place nearly ten years earlier, the *EM* Editor took steps to stop an escalation: "We insert this letter with reluctance; but Mr. Holmes quite unnecessarily provoked it by his letter…. It is, and always has been, a matter of grief to us, as we are sure it is to many readers, that these bickerings have wasted our space, and cost us once or twice the help of highly esteemed and most valuable helpers. We will not insert or take any notice of any rejoinder from anybody on this occasion.—Ed."

Whilst there was no immediate rejoinder, the dispute between Hunter and Holmes rumbled on throughout 1907. Hunter wrote: "I find myself compelled to state that..….Mr. Holmes has now reached his third charge of dishonesty against me, since about ten years ago I began to write in these pages under the penname I now use. He has been repeatedly, during that interval, shown his errors in so doing, by others as well as by myself, and yet he has never once deigned to tender an apology or make the slightest excuse for that conduct". (55)

In the following edition of the *EM*, matters were becoming too much for Holmes who referred to various prominent astronomers with whom he had disagreed: "I have yet to learn that to be unable to accept a man's doctrines is to attack him (or to point out a trivial error either). I am aware Mr. Maunder, for whom I have every respect, treated my queries re Solar disturbances as an attack, and I regretted it. Mr. Sadler regarded want of belief in his story as personal; but he was only repeating what he was told, and so was too impulsive. If I have offended Mr. Chambers (56), I am sorry….. It appears I have many who object to my membership of the B.A.A. That being so, it is in the power of the council to expel me, and it is quite possible I shall resign my membership at the end of the present session, and so remove one stumbling-block to the progress of the B.A.A." (57)

The *EM* Editor was clearly exasperated as he testily appended at the bottom Holmes's letter: "This is the last letter we will insert from anybody in regard to this squabble, of which we are sick. It is, as usual, accompanied by a private letter from Mr. Holmes accusing us of unfairness and partiality to 'H.,' because we will not insert Mr. Holmes's letters verbatim. We can only say (as 'H.' knows perfectly well) we have done our best to omit what seemed to us irrelevant and mere personal allusions from the letters of both principal disputants, for whom we have much esteem, and whose many past contributions we have valued, coupled with many regrets that bitterness mars them occasionally. - Ed."

In spite of Holmes's threatened resignation, he continued to attend most BAA meetings and wrote in the *Journal* for several more years. He did, however, stop writing to the *EM* for several months, prompting the US subscriber, Daniel W. Edgecomb (1840-1915), to write "I would really like to say that I hope Mr. Edwin Holmes will long continue his observations, both telescopic and philosophic, herein printed. There is at least one long-distance reader who likes to see a whole column or more with his name at the bottom" (58).

One of Holmes's first letters to the *EM* on resuming his correspondence in 1908 August he reviewed various theories on the origin of the lunar ray systems (59). In a side-swipe at the Editor of the BAA *Journal*, F.W. Levander (1839-1916), Holmes said he would not send his ideas to the *Journal* as they might not meet the "approval of the censor". And he inserted a further slur against the BAA in a letter shortly afterwards: "I am sorry to see the British Astronomical Association has gone further on the down grade, and numbers more than 100 less than a year ago. The causes are not far to seek, but I may not enter on them here; but my expression of regret may, I hope, pass." (60)

Several years later Holmes still believed that the Editor of the *Journal,* still F.W. Levander, did not welcome his contributions: "It may be asked why I do not send this [his views on meteor observations] to the British Astronomical Association. There are several reasons; but one is sufficient. Such papers are unwelcome there, and have to pass a censorship unless they are written by one of the élite" (61).

**The Canals of Mars**

Towards the end of the nineteenth century canal fever had broken out in the astronomical world with increasing numbers of observers reporting that they had seen Giovanni Schiaparelli's (1835-1910) *canali* on the Red Planet. One of the earliest people to confirm the presence of canals was Camille Flammarion (1842-1925) followed by two of their greatest proponents: Percival Lowell (1855-1916) and W.H. Pickering (1858-1938). Others were more sceptical, with many claiming they simply did not exist, including Nathaniel E. Green (1823-99) and E.W. Maunder. Needless to say Holmes had a strong opinion on the matter, which was that the Martian canals were figments of the imagination. This led to a debate with Flammarion in the pages of the *EM*, with Holmes criticising Flammarion's journal, *L'Astronomie*, for printing so many observations of the canals: "These lines are straight in most instances, but the most marvellous thing about them is that *they remain straight in all positions on the disc of the planet*. Is not this fatal to any notion of their objective reality! A straight line on the meridian of Mars would necessarily become an apparent curve when near the border of the disc. The lines supposed to be seen do not so become curved. What is the legitimate inference?" (62). He was no less dismissive of Flammarion's view that the canals had water coursing through them: "Mr. Flammarion regards these 'canals' as rivers. Now as the surface of Mars is cut up into mincemeat by the network of lines, I think one may fairly ask how, where, and by what means a few dozen rivers to which the Amazon and Mississippi are driblets from a teapot, can possibly rise and run their course on such a small globe as Mars?"

So well known was Holmes's antagonism towards the canals that when Arthur Mee sent him a drawing of the Red Planet that he had made with his 8½-inch (22 cm) reflector on 1897 January 4 that he humorously annotated it with the words: "Please excuse the absence of canals"! (63)

As if canals on Mars were not enough, observers began to report similar markings on other planets, prompting Holmes to put into words his abhorrence: "The canal disease is spreading. Only Mars was first affected, then Venus, and now Jupiter is beginning to break out. One can only hope our poor old earth will escape the scourge". (64)

Holmes's next argument about the Martian canals was with Eugène Antoniadi (1870-1944), Director of the BAA Mars section and a celebrated visual observer of the planet. Antoniadi was employed by Flammarion as an observer at his Juvisy observatory between 1893 and 1902. While Antoniadi never doubted that Mars was habitable and likely inhabited, his initial views of whether canals were present were more equivocal (65). However, as the 1890s progressed he grew increasingly critical of the ever more complex canal systems being reported by Lowell, Pickering and others, becoming more certain in his opinion that they were due to optical illusion. This conclusion was crystallised by his own observations made towards the end of 1909 using the *Grand Lunette* at the Meudon Observatory, which provided the clearest views of Mars he had ever had – and it was plain to see that no canals were present whatsoever. He began to communicate his observations through a number of publications and there was much discussion in the pages of the *EM*.

Holmes was naturally delighted that the canals had finally been debunked: "Eighteen or twenty years ago the few of us who disputed or denied the existence of the geometrical network on Mars were regarded as little better than blasphemers. There has been an entire change in the interval, and now allusions to this network are received with derision, and the people who were most resentful at the propounding of doubts are in many cases taking the credit of opposition, and, in some cases, both seeing and denying, in almost the same breath" (66). In a further letter he mentioned "how very uncertain M. Antoniadi's attitude has been on the canal question" (67). This clearly vexed Antoniadi, who retorted: "Mr. Holmes obviously misunderstands my position in the 'canals' of Mars, as he ventures to insinuate that I believe and disbelieve at the same time in the reality of these markings…. [he] seems to exult in what he calls the uncertainty of my past attitude in the 'canal' question. The truth is that I always felt sceptical on the reality of the geometrical network" (68). The argument between Holmes and Antoniadi went back and forth over the next few weeks, with increasing animosity, each accusing the other of bad faith and deliberately misrepresenting the others' views. Holmes's parting comment in the exchange was that he did "not intend to write again on this matter, and, indeed, should not have done so now but for his [Antoniadi's] charges of dishonest practices and motives, which I could hardly pass over without explanation" (69). Antoniadi, clearly offended and exasperated by Holmes's criticisms, also chose to let the matter rest at that point.

**Declining health**

By 1914, at the age of 75, Holmes's health was beginning to decline. He had not been able to walk for 6 months, although his eyesight was still excellent (70) and he was still able to observe, mainly with a 3-inch (7.5 cm) refractor, with which he followed the 1914 November transit of Mercury (71). At about this time he became engaged in a long-running dispute with W.F.A. Ellison on the merits of the Foucault test for telescope mirrors versus star tests. The Reverend William Frederick Archdall Ellison (1864-1930; Figure 9), a BAA member and regular contributor to its *Journal* and the *EM*, was well-known for making high quality optics, would go on to become Director of the Armagh Observatory from 1918 (72). Ellison himself also had quite a reputation for using forthright language in his letters. As had become his style, Holmes took a combative position. He objected that Ellison was quoting him out of context, a claim with Ellison also made against Holmes. Clearly this was not going well and Holmes felt compelled to continue to take the debate further in spite of his health worries: "I am not well enough to write more, and should not have written this but for Mr. Ellison so continually misrepresenting my position. Of course, it distracts attention from his own misstatements, and is a sort of carrying the war into the enemy's country" (73). Although the *EM* Editor tried to draw a line with the familiar insertion of "This topic has had all the space we can spare—ED.", it rumbled on for several more weeks.

Holmes's last letters to the *EM* appeared in the edition of 1915 June 4; one was on "Cutting holes in gears" (74) and another on microscope objectives (75). Whether the fight had simply gone out of him as a result of poor health we do not know for sure. He didn't retire until 1918 (76) and he still maintained a personal correspondence with other astronomers until early 1919 when his health finally faded quickly until his death in November (77). A brief note on Holmes's passing was written in the *Journal* by the Comet Section Director, A.C.D. Crommelin (1865-1939) (76) and a short obituary by Arthur Mee (6).

**Holmes the man**

Something of Holmes's character emerges through his vast correspondence in the *EM*, spanning nearly five decades and including more than 650 contributions; in some editions there were six separate items. As we have seen from the analysis of some of the debates and disputes he became involved with, he could be robust and opinionated when setting out his views. He could also be infuriating and offensive; and he clearly had the knack of rubbing people up the wrong way. A frequent tactic was to accuse his opponents of twisting his words, something which he was not averse to doing himself, putting words in people's mouths which they did not intend to say. He frequently alleged bad faith in people's motives, an example being his lengthy debate with Antoniadi over the canals of Mars. Whether he deliberately did this to cause annoyance or simply to provoke a debate is not always clear. On some occasions he clearly caused great offence, for example his correspondence with Alfred Allen of Sheffield, which he not only conducted in the pages of the *EM*, but also through private correspondence. Only one person ever truly got the measure of him and managed to give as good as he got: James Hunter. Hunter's use of humour and good-natured teasing, as exemplified by his linking of Edwin Holmes the astronomer with his namesake Sherlock Holmes, the great, but deceased detective would doubtless have irritated him. Hunter shared Holmes's characteristic persistence: whilst Holmes might have worn down other disputants to allow him seize the last word on a matter, Hunter was made of equally stern stuff and would not give up.

As we have seen there were several occasions when the Editor of the *EM* had to draw a line under a debate. Usually Holmes would comply with the Editor's edict, sometimes ceasing further correspondence for months. However, later in life even Editorial criticism did not deter him as he strove to seize the last word in a debate. So why did not the Editor simply stop publishing his letters? Mainly because much of what Holmes wrote was helpful, insightful and interesting, if only the personal aspects could be removed. Many people appreciated reading his contributions, finding them instructive – and some would write in to say how much they missed him when he stopped writing for periods of time. Even some of those who were subjected to Holmes's attacks, such as Arthur Mee, became good friends and respected him for his knowledge. On the other hand, others simply could not tolerate

Holmes's approach. For example Robert Barker (1873-1966), a prominent member of the BAA Lunar section (78), said "I certainly took no delight in Mr. Edwin Holmes' letters; they were often indited with a vitriolic pen" (79).

We have also seen how Holmes became a critic of the BAA, its organisation and its Officers, notably the *Journal* Editor. This was exacerbated by the major public disagreement he had with E.W. Maunder, who for many *was* the BAA. Holmes disliked the way, as he saw it, that some of the more experienced and well-known members remained aloof from the general members and cited how some members did not feel welcome at meetings, fearing to ask questions lest their ignorance be exposed by the heavyweights. On the other hand Holmes had shown great support for the Association over a period of many years from its beginnings in 1890. He published many papers in the *Journal* and was a regular contributor to its meetings, rarely missing one. Certainly nobody was immune from Holmes's criticism, even if they were part of the astronomical establishment, as can be seen from the many famous names with whom he became entangled. In fact, it was almost as if Holmes deliberately wanted to take on the establishment and those whom he considered to be self-important. He also saw himself as the self-appointed voice of the silent majority of members who, he claimed, were not fully catered for by Association.

It was not only in connexion with astronomy that these disputes arose. There were similar protracted debates with members of the microscopy community, where Holmes again frequently employed the tactic of painting himself as the maligned party. During a heated debate on the design of microscope objectives he claimed "I expect I shall be misquoted and misrepresented, as usual, so I may as well say I have no intention of noticing comments or entering on any disputes, being quite sure that later on someone will adopt my views as their own, and assert I wrote something quite different" (80).

Although the vast majority of Holmes's letters were in connexion with astronomy, microscopy and optics, he held strong views on other subjects such as politics and health matters. For example he was against mass vaccination against smallpox as he thought it an infringement of civil liberty; he also wrote on alcohol consumption (acceptable in moderation), factors influencing a person's susceptibility to tuberculosis, and the benefits of a balanced diet. He was also interested in hypnotism and auto-suggestion.

But we should avoid judging a man solely on his writings. The rancour associated with Holmes's letters does not tell the whole story of Holmes's character. Certainly he could be stubborn, opinionated and vitriolic in his writing. He could also be precise to the point of pedantry. H.P. Hollis probably got the balance right when he referred to "the trenchant letters of Mr. Edwin Holmes", whilst going on to concede that "Mr. Holmes was a thoughtful writer, and a keen observer" (81). There were also other factors that influenced his writings, notably during his wife's illness which led to her death in 1907. Shortly after her passing, amidst a wave of condolences from *EM* readers, Holmes acknowledged that "Many of my letters of the last three years were written in order to distract my own attention from troubles" (82). On the other hand, it seems that many of those who knew him in person saw a different side of his character in real life from the one that emerges from his letters and he had many positive qualities. He was knowledgeable about many aspects of astronomy and he had much practical experience of optics and observing which he took time to share with others, especially those less experienced. Above all he was enthusiastic and engaging. As a result many people were grateful for his advice, especially novices, and held him in high regard.

One person who had an especially high regard for Holmes was the Reverend T.H.E.C. Espin (1858-1934) who maintained a correspondence with him for many years and knew him in person – Espin's photograph of Holmes appears as Figure 1. Holmes had reported his observations made with his 12¼-inch reflector of some double stars which he thought had not previously been recorded in several letters to the *EM* in 1901 and 1902. After Holmes's death, Espin took it upon himself to investigate these further and he found it "a congenial task to rescue these stars" (81) as a tribute to his old friend. Espin's job was not helped by the fact that Holmes had observed with an altazimuth mount, so his reported positions were often inaccurate; moreover it appears that Holmes did not keep an original notebook from which to corroborate his observations. Nevertheless Espin re-measured the stars and published a list of 41 stars in the *Monthly Notices* of the RAS in 1926 entitled *The Late Mr. Holmes' Double Stars* (83). He enlisted the assistance of the young Mervyn Archdall Ellison (1909-1963), son of the Reverend W.F.A. Ellison with whom Holmes had had several disputes; Ellison Junior had access to the 10-inch (25 cm) Grubb refractor and 18-inch (46 cm) Calver reflector at Armagh, where his father was Observatory Director. A page from Mervyn Ellison's observing notebook is shown in Figure 10.

Thus two sides of Holmes's character emerge: the controversial Holmes we see in his letters and the genial man we learn about from people who knew him in person. It is, of course, not uncommon for people to come across differently in their writings from how they are in real life.

In spite of having a long career as an amateur astronomer, and making an important contribution to the BAA in its early years, Holmes might have slipped into obscurity had it not been for the chance discovery of his comet in 1892. Whenever Comet 17P returns to the sun's vicinity in its short orbit it will be carefully tracked by telescopes in case it once again undergoes a bright outburst similar to the one that brought it to Holmes's attention. And when it does, the world will be reminded of Edwin Alfred Holmes.

**Acknowledgements**

I am most grateful for the assistance I have received from a great many people whilst preparing this paper. Adam Perkins, Scientific Manuscripts Collections, Department of Manuscripts & University Archives, University Library, Cambridge kindly arranged access to some papers relating to Holmes held in the RGO archives held at the University of Cambridge Library. Commander William F.A. Ellison, RN (retired), son of Mervyn Ellison and Grandson of the Reverenced W.F.A. Ellison provided copies of his late father's observing notebooks. Several people gave me permission to use images: Professor Mark E. Bailey of the Armagh Observatory, the photograph of the Reverend William Frederick Archdall Ellison; Michele and Peter Clare, the photograph of Alfred Henry Allen; Mark Hurn, Departmental Librarian, University of Cambridge, Institute of Astronomy, Newall's 1892 sketch of 17P/Holmes; and Bill Becker, the photograph of the Reverend Charles L. Tweedale. Dr. Richard Miles allowed me to reproduce his splendid CCD image of the 2007 outburst of 17P. Vicki Hammond, Journals & Archive Officer of The Royal Society of Edinburgh/RSE Scotland Foundation provided biographical details of James Hunter. Mike Saladyga, AAVSO, gave insights into the life of D.W. Edgecomb. I wish to record my sincere thanks to all these people.

During my research I made extensive use of scanned back numbers of the BAA Journal, which exist largely thanks to the efforts of Sheridan Williams, as well as of the English Mechanic, thanks to Eric Hutton. These are truly wonderful resources for historians of nineteenth and twentieth century British astronomy. I also used the NASA/Smithsonian Astrophysics Data System.

Finally, I thank the referees, Mike Frost and Denis Buczynski for their helpful comments.

**Address**: "Pemberton", School Lane, Bunbury, Tarporley, Cheshire, CW6 9NR, UK

**References and notes**

1. The new star, S And, was discovered on 1885 August 19 by Isaac Ward (1834-1916) of Belfast. It was the first ever supernova to be discovered outside the Milky Way.

2. Holmes E., Observatory, 15, 441-443 (1892).

3. The comet was independently discovered on the evening of 1892 November 8 by Dr. Thomas David Anderson (1853 –1932) of Edinburgh, who, only 9 months earlier had shot to fame through his discovery of Nova Aurigae 1891. An independent discovery was also made by J. E. Davidson, at Mackay, Queensland on November 9.

4. The comet was observed during its 1899 and 1906 returns, but was lost for the next seven returns until 1964. It has been observed at each return since. Its orbit currently lies between 2.05 and 5.2 AU from the sun. It is subject to perturbations by Jupiter.

5. Mee was an original member of the BAA. He also set up the Astronomical Society of Wales, which for a time drew a national membership. Holmes is listed as an Associate Member in 1898, 1905 and 1910. He is not to be confused with Arthur Henry Mee (1875 – 1943), who was a British writer, journalist and educator.

6. Mee A., JBAA, 29, 113 (1919).

7. Holmes described the construction of the observatory, which cost about £3, in Holmes E., JBAA, 14, 283-284 (1904). One problem with the observatory was that it rather small for the size of telescope deployed and the shutter only allowed a restricted view of the sky. The view was further curtailed by the proximity of the house.

8. Holmes bought the telescope from a Mr Upton.

9. It appears that Holmes had at least three 12¼-inch mirrors at various times. One made by himself, one made by George With and the other by J. Linscott (who purchased With's mirror making tools when With ceased work). Holmes seconded Linscott's (who was described as "Optician, Ramsgate") BAA membership application in 1893.

10. Holmes E., English Mechanic, 302, 373 (1871).

11. Holmes E., English Mechanic, 493, 652 (1874).

12. Hampden J., English Mechanic, 495, 19 (1874).

13. Holmes E., English Mechanic, 497, 75 (1874).

14. Hampden was so certain that the Earth was flat that in 1870 he offered £100 to anyone who could prove it was round. Alfred Russel Wallace accepted the offer and performed what was known as the Bedford Level Experiment. Wallace won the wager and took the money. Hampden later took Wallace to court and legal proceeds went back and forth. Hampden tried to get his views published in the EM, but the Editor refused to print most of his letters.

15. Allen was one of the first Public Analysts in the United Kingdom and the first appointed by the City of Sheffield. He was a founding member of the Society of Public Analysts and of the Institute of Chemistry. Allen's biography appears in Clare P. and Clare M., Journal of the Association of Public Analysts (Online), 40, 39-59 (2012).

16. Allen A.H., English Mechanic, 789, 211 (1880).

17. Holmes E., English Mechanic, 932, 498 (1883).

18. A biography of Franks will appear in: Shears J., JBAA, accepted for publication (2013).

19. Franks W.S., English Mechanic, 1075, 177 (1885).

20. Holmes E., English Mechanic, 1079, 260 (1885).

21. Franks W.S., English Mechanic, 1080, 282 (1885).

22. Calver G., English Mechanic, 1080, 282 (1885).

23. Holmes E., English Mechanic, 1082, 318 (1885).

24. Quote from Noble's RAS obituary: MNRAS, 65, 324-343 (1905).

25. Noble's obituary (Hollis H.P., The Observatory, 27, 298-300 (1904)) mentions "Each fortnight, for many years past, a long letter has appeared in the English Mechanic over the signature "A Fellow of the Royal Astronomical Society", which it is an open secret was a pseudonym for Capt. Noble, often containing criticisms of the Kensington Science Department, almost libellous, but those who knew the writer were aware that 'his bark was worse than his bite'".

26. Holmes E., English Mechanic, 1489, 154 (1893).

27. Holmes E., English Mechanic, 1505, 510 (1894).

28. Mee A., English Mechanic, 1506, 534 (1894).

29. Letter to English Mechanic, 1511, 61 (1894).

30. Holmes E., English Mechanic, 1513, 107 (1894).

31. One of Hunter's proposers when he came to be elected to the Royal Society of Edinburgh was Charles Piazzi Smyth (1819-1900), one time Astronomer Royal for Scotland. Hunter died in Edinburgh in 1921.

32. Conan Doyle had killed off Sherlock Holmes in "The Final Problem", published in 1893.

33. English Mechanic, 1728, 267 (1898).

34. "A.S.L." was posthumously disclosed as a Mr Lukin, see Hunter J., English Mechanic, 2385, 434 (1910).

35. Hunter J., English Mechanic, 1753, 257 (1898).

36. Holmes E., English Mechanic, 1753, 258 (1898).

37. Holmes E., English Mechanic, 1877, 97 (1901).

38. English Mechanic, 2008, 145 (1903).

39. Holmes E., JBAA, 1, 215 (1891).

40. Holmes E., JBAA, 27, 121 (1917).

41. Holmes's first recorded comment in the Journal was in the report of the meeting of 1891 January 28.

42. During the BAA meeting of 1894 March 28, Holmes did make some light-hearted criticism of the accuracy of some drawings made by Professor James Keeler (1857-190) at the Lick Observatory. These were subsequently reported in the Journal and were evidently read by Keeler who wrote to the Journal to object and to explain his drawings. See: Keeler J.E., JBAA, 4, 358 (1894). A somewhat taken aback Holmes commented that "he had no intention of hurting Prof. Keeler's feelings at all".

43. Holmes E., JBAA, 16, 272 (1906). "Notes and Queries Incidental to Mr. Maunder's Researches".

44. Hollis H.P., English Mechanic, 2172, 327 (1906).

45. Buss A.A., English Mechanic, 2175, 403 (1906).

46. Grover C., English Mechanic, 2175, 403 (1906).

47. Holmes E., English Mechanic, 2176, 425 (1906).

48. Buss A.A., English Mechanic, 2177, 451 (1906).

49. S.S., English Mechanic, 2178, 472 (1906).

50. English Mechanic, 2179, 500 (1906).

51. Holmes E., English Mechanic, 2181, 544 (1907).

52. See report of the BAA of 1906 June in JBAA, 16, 331 (1906).

53. Holmes E., English Mechanic, 2190, 137 (1907).

54. Hunter J., English Mechanic, 2192, 188 (1907).

55. Hunter J., English Mechanic, 2229, 427 (1907).

56. G.F. Chambers (1841-1915): BAA member and author of popular astronomy books.

57. Holmes E., English Mechanic, 2230, 451 (1907).

58. English Mechanic, 2239, 60 (1908). Edgecomb was a BAA member and lived at Fairfield, Connecticut. His letter was largely claiming the superiority of reflectors over refractors. This was of course a well known debate, prompting the EM Editor to write: "To save trouble, readers will please note we shall not insert any comments or replies to this letter. It comes from an old and esteemed correspondent a long way off who has not favoured us now for several years". Edgecomb was by profession a lawyer. He was also something of a polymath. He was an inventor and telescope maker. He oversaw the construction of the first electric street railway car in New York. He served as Secretary of State in Connecticut in 1873.

59. Holmes E., English Mechanic, 2265, 64 (1908).

60. Holmes E., English Mechanic, 2270, 183 (1908).

61. Holmes E., English Mechanic, 2506, 221 (1913).

62. Holmes E., English Mechanic, 1260, 238 (1889). The italics in the quote are Holmes's.

63. Mee's hand-coloured sketch appeared on a postcard he sent to Holmes. The postcard is preserved in the University of Cambridge Library, ref RGO 45/40.

64. Holmes E., English Mechanic, 1685, 483 (1897).

65. Antoniadi's life, observational work and his evolving view on the Martian canals are admirably described in two papers by Dr. Richard McKim: McKim R., JBAA, 103,164-170 (1993) and McKim R., JBAA, 103,219-227 (1993).

66. Holmes E., English Mechanic, 2337, 540 (1910).

67. Holmes E., English Mechanic, 2338, 562 (1910).

68. Antoniadi E., English Mechanic, 2339, 584 (1910).

69. Holmes E., English Mechanic, 2343, 62 (1910).

70. Holmes E., English Mechanic, 2562, 305 (1914).

71. Holmes E., English Mechanic, 2590, 342 (1914).

72. Ellison's book "The Amateur's Telescope" (1920) was considered a standard for telescope-makers for many years.

73. Holmes E., English Mechanic, 2592, 386 (1914).

74. Holmes E., English Mechanic, 2619, 403 (1915).

75. Holmes E., English Mechanic, 2619, 401 (1915).

76. Crommelin A.C.D., JBAA, 29, 84 (1919).

77. Holmes's death was announced in a letter to the EM by R.J. Tait of Chester-le-Street. Tait R.J., English Mechanic, 2810, 21 (1919).

78. Robert Barker established a lunar observing group within the BAA in 1930, known as "Barker's Circle", see McKim R., JBAA, 123, 20-32 (2013).

79. Barker R., English Mechanic, 3138, 261 (1925).

80. Holmes E., English Mechanic, 2612, 247 (1915).

81. Hollis H.P., English Mechanic, 3175, 37 (1926).

82. Holmes E., English Mechanic, 2220, 221 (1907).

83. Espin T.H.E.C., MNRAS, 86, 79-80 (1926).

84. Papers of John Guy Porter: photographs and papers of E.A. Holmes. Reference mark RGO 45/40. These papers were given to Porter in 1950 by Holmes's daughter-in-law "Mrs Holmes".

85. Photograph inscribed on verso: "The Rev. Charles L. Tweedale and Mrs. Tweedale with the spirit - form of the late F. Burnett who died in 1913. Taken under good test conditions Sept. 5th 1919." The "extra" shown is that of Rev. Tweedale's father-in-law, who was supposedly never photographed in life with a beard except when wearing his hat. That fact was considered evidence of the supernatural origin of this image. The photographer, William Hope, began taking spirit photographs in 1905. Hope was part of the "Crewe Circle" the members of which were known for their spirit photography. Hope was exposed as a fraud in 1922.

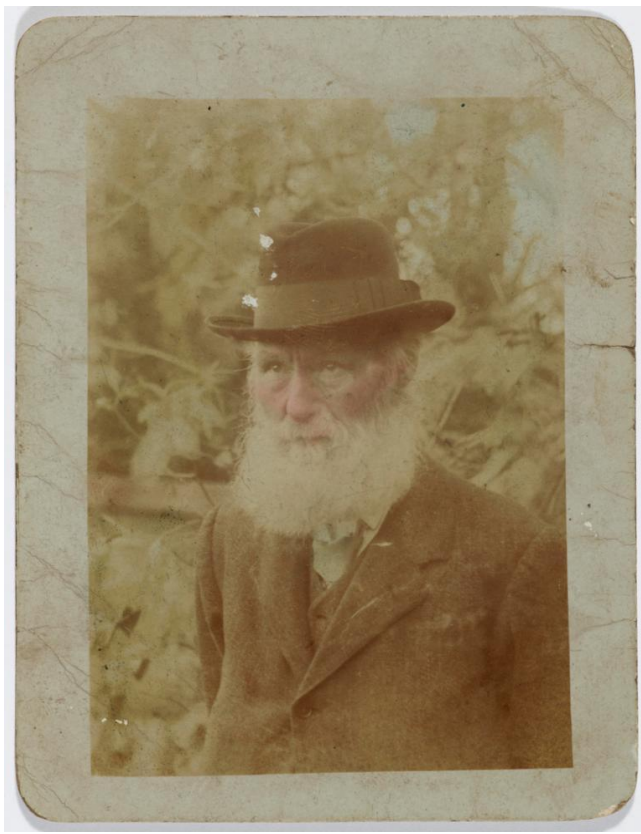

Figure 1: Edwin Holmes (photograph by T.H.E.C. Espin). Courtesy of University of Cambridge Library (84)

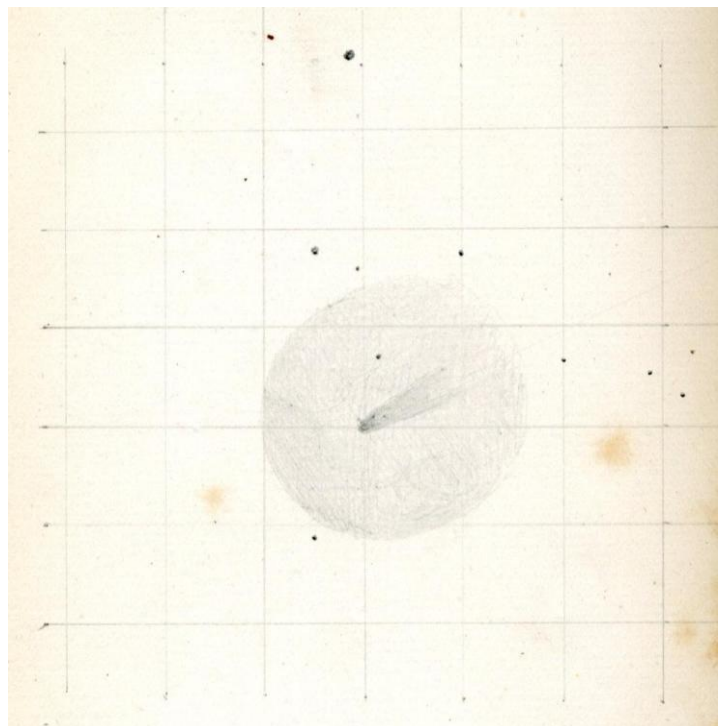

Figure 2: 17P/Holmes in outburst on 1892 November 14. Sketch by H.F. Newall, who annotated it: "Intention of sketch is to show clear boundary on proceeding side and hazier on f[ollowing] side". Courtesy of University of Cambridge, Institute of Astronomy Library.

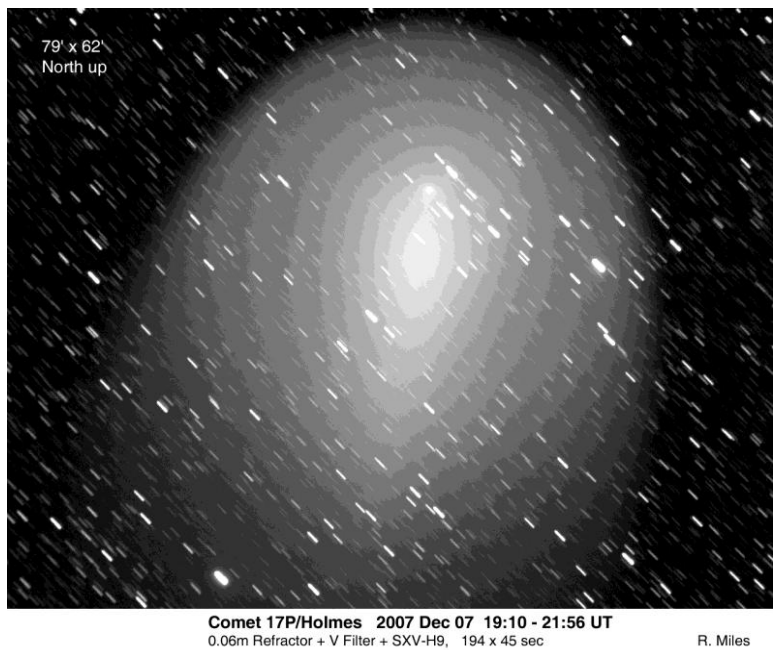

Figure 3: 17P/Holmes in outburst on 2007 Dec 7 (Richard Miles)

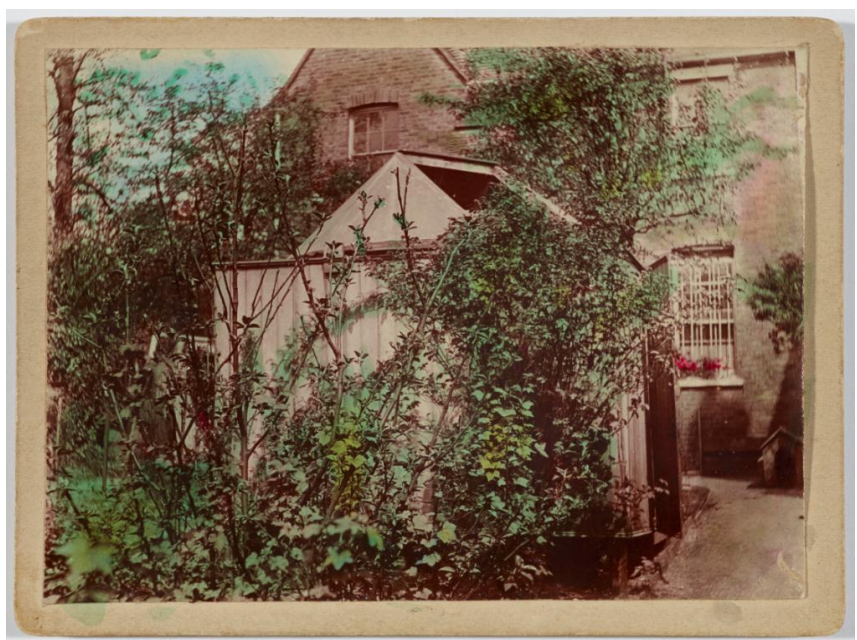

Figure 4: Holmes's observatory at Hornsey Rise, London, 1901 May 3. Courtesy of University of Cambridge Library

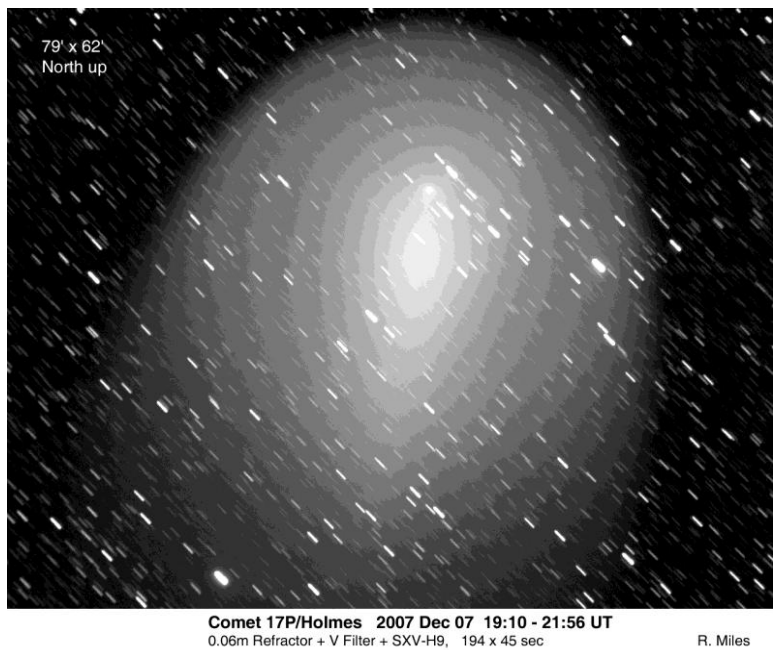

Figure 3: 17P/Holmes in outburst on 2007 Dec 7 (Richard Miles)

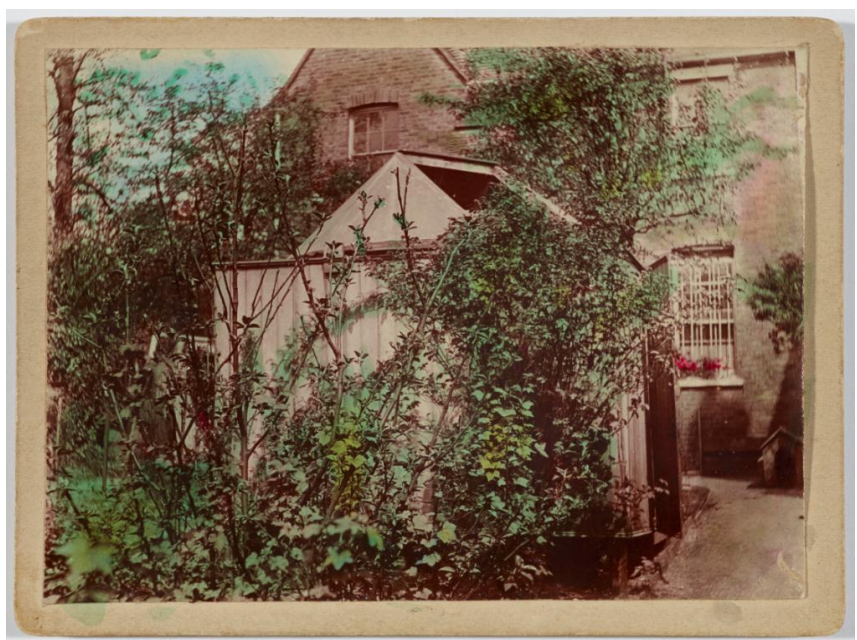

Figure 4: Holmes's observatory at Hornsey Rise, London, 1901 May 3. Courtesy of University of Cambridge Library

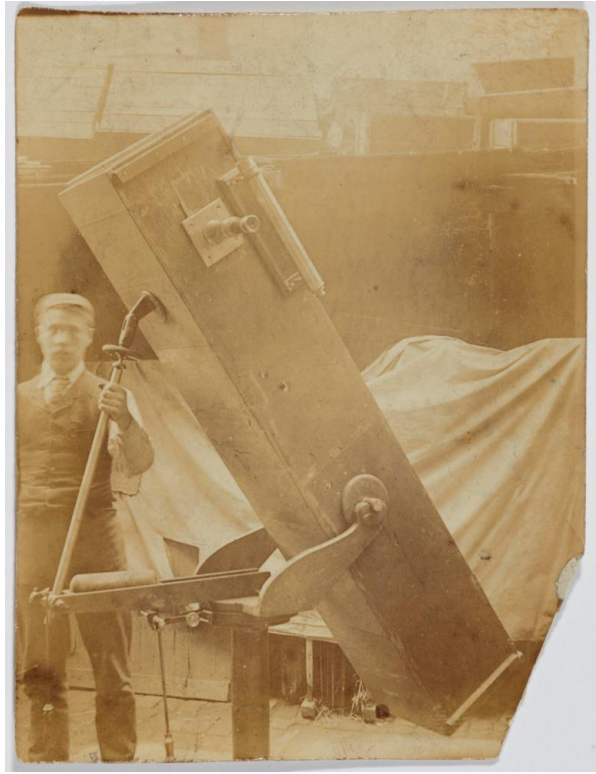

Figure 5: Holmes's 9-inch Newtonian reflector. The identity of the person standing by the telescope is unknown. Courtesy of University of Cambridge Library

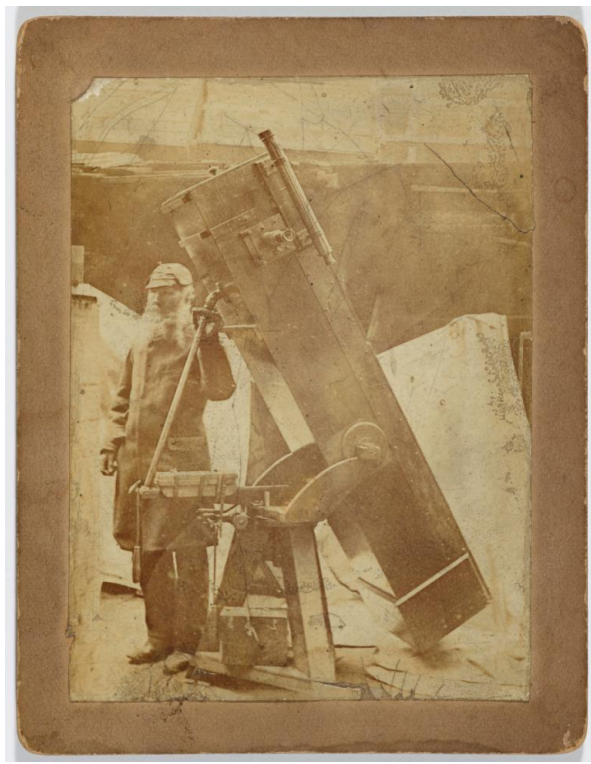

Figure 6: Holmes and his 12¼-inch Newtonian reflector, possibly the one with which he discovered 17P/Holmes. Courtesy of University of Cambridge Library

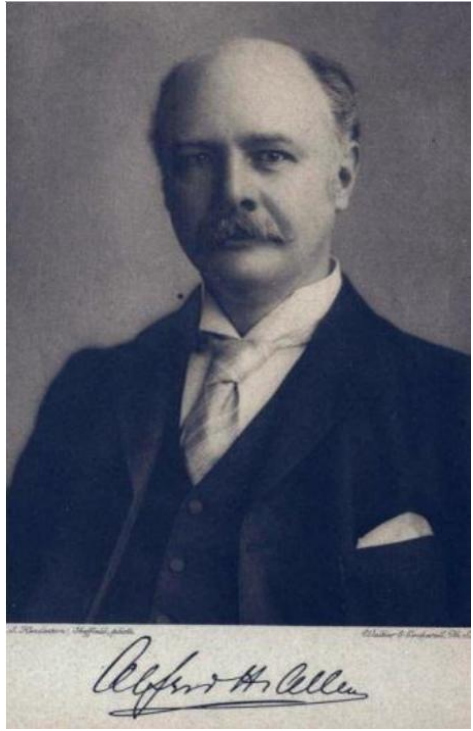

Figure 7: Alfred Henry Allen (1846 – 1904). Courtesy of Peter and Michele Clare

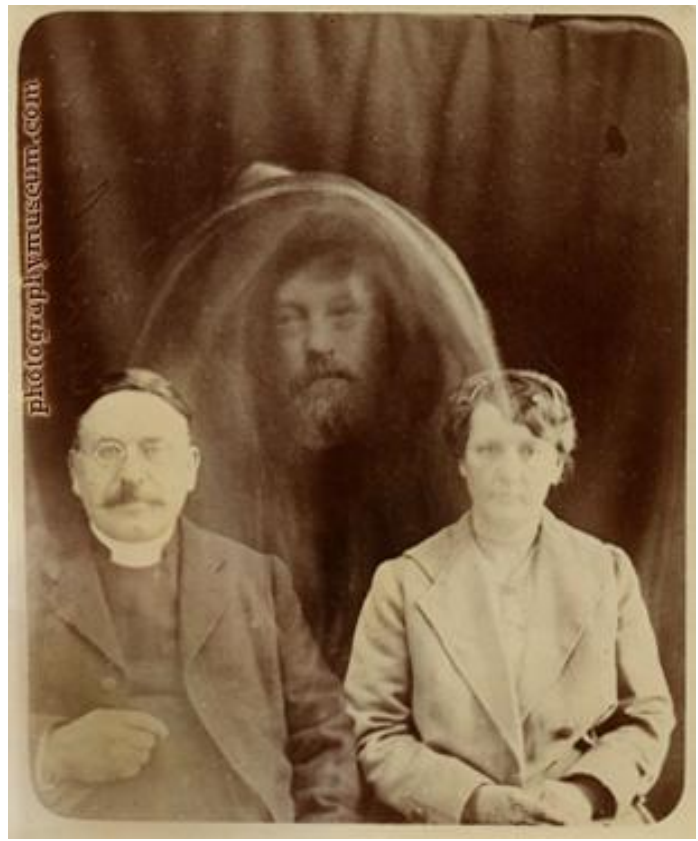

Figure 8: The Reverend Charles L. Tweedale, FRAS, and Mrs. Tweedale, supposedly with the spirit form of his late father-in-law F. Burnett in 1919. Photograph taken by William Hope (1863-1933) of Crewe (85). Copyright Wm. B. Becker Collection/American Museum of Photography

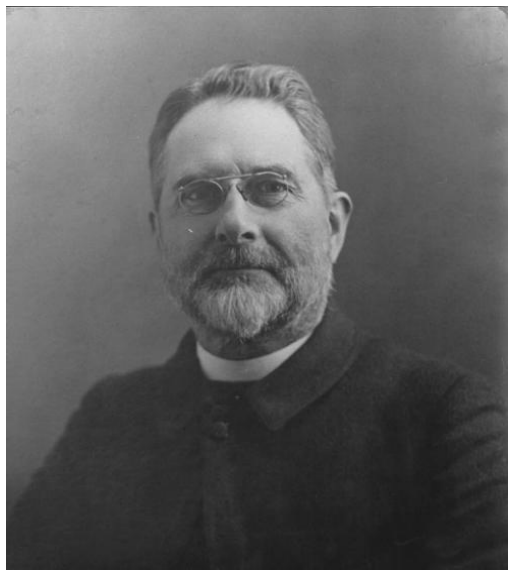

Figure 9: The Reverend William Frederick Archdall Ellison (1864-1930), sixth Director of Armagh Observatory. Image copyright Armagh Observatory

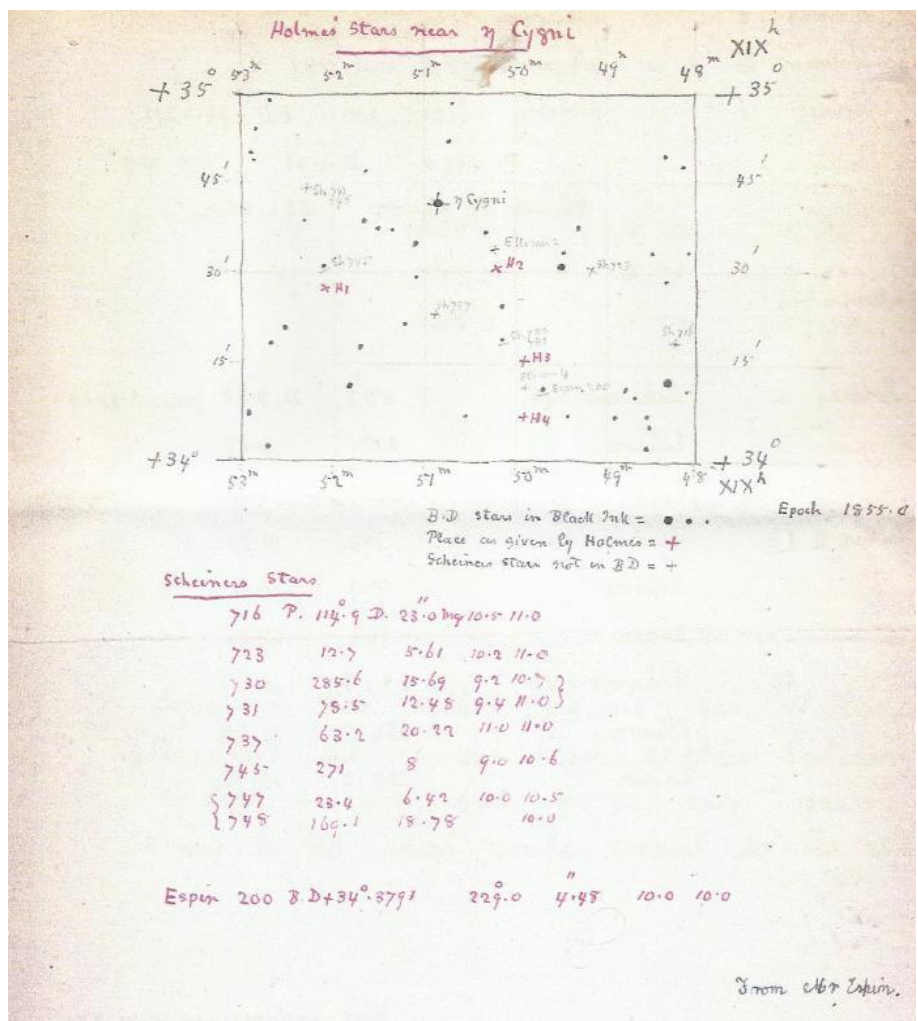

Figure 10: A page from the notebook of Mervyn Ellison (1909-1963) showing some of Holmes's double stars (marked +) near η Cyg. Courtesy of Commander W.F.A. Ellison